\documentclass[aps,prd,10pt,
               reprint,
               notitlepage,
               preprintnumbers,numbers,sort&compress,nofootinbib,
               showpacs, 
               colorlinks,
               linkcolor=blue,
               citecolor=blue]{revtex4-1}
\usepackage{graphicx,amsmath,amssymb,bm}
\usepackage{psfrag}
\usepackage{feynmp}
\usepackage{hyperref}
\usepackage{enumitem}

\usepackage{microtype}

\newcommand{\exclude}[1]{}

\usepackage{microtype}
\newcommand{\nn}{\nonumber}
\newcommand{\beq}{\begin{equation}}
\newcommand{\eeq}{\end{equation}}
\newcommand{\be}{\begin{eqnarray}}
\newcommand{\ee}{\end{eqnarray}}

\def\dd{ \,\mathrm{d} }

\def\+{\dagger}
\def\la{\langle}
\def\ra{\rangle}
\def\<{\langle}
\def\>{\rangle}

\newcommand{\Lqcd}{\Lambda_{\mathrm{QCD}}}

\begin{document}

\title{ Entropy,   Contact Interaction with    Horizon and    Dark Energy}

\author{ Ariel R. Zhitnitsky}
\affiliation{Department of Physics \& Astronomy, University of British Columbia, Vancouver, B.C. V6T 1Z1, Canada}

\begin{abstract}
We present some arguments suggesting that the mismatch between  Bekenstein- Hawking entropy and the entropy  of entanglement for vector fields   is due to the same gauge configurations which saturate the contact term in topological susceptibility
in QCD.
In both cases the extra term with a ``wrong sign" is due to distinct  topological sectors in gauge theories. This extra term 
has   non-dispersive nature, can not be restored from conventional spectral function through dispersion relations, 
and  can not be associated  with any physical  propagating degrees of freedom. We make few comments on some profound consequences of our findings. In particular,   we speculate that the source of the observed dark energy may also be related to the same type of gauge configurations which are responsible for the mismatch between black hole entropy and the entropy of entanglement
 in the presence of  causal  horizon.

 \end{abstract}
\maketitle

  \section{Introduction}
  The relation between black hole entropy and entropy of entanglement for matter fields has been a subject of intense discussions  
 for the last couple of  years, see reviews\cite{Jacobson:2003vx,
Fursaev:2004qz,Frolov:1998vs,Solodukhin:2011gn}  and original references therein. There are many subtleties in relating these two things. The present work is concentrated just on one specific subtlety first discussed in 
\cite{Kabat:1995eq}. Namely, it has been claimed  \cite{Kabat:1995eq} that for spins zero and one-half fields the one loop correction to the black hole entropy is equal to the entropy  of entanglement while for spin one field the black hole entropy has an extra term describing the contact interaction with the horizon. Precisely this  contact interaction with the horizon is the main topic of the present work. Before we elaborate on this subject we want to make one preliminary remark  regarding the  term ``black hole entropy". As it has been argued in a number of papers, see e.g.\cite{Kabat:1994vj,Callan:1994py,Jacobson:1999mi} and  also review papers~\cite{Jacobson:2003vx,Fursaev:2004qz,Frolov:1998vs,Solodukhin:2011gn}  the notion of black hole entropy should apply not just to black holes but to any causal horizon (``black hole entropy without black holes").  We adopt this viewpoint, and in fact we shall not discuss black hole physics in this paper at all. Rather, the main application of our studies will be cosmology of the expanding  universe
   and its causal horizon. For short, we shall use    term ``entropy" through the paper. 
  
  The unique features of the contact  term related to the vector gauge field in the entropy computations
can  be summarized as follows~\cite{Kabat:1995eq} (see also follow up paper \cite{Iellici:1996gv}) :
\begin{enumerate}[label=\alph*)]
\item the contact term being a total derivative can be represented as a surface term determined by the behaviour at the horizon;
\item this term makes a {\it negative}  contribution to the black hole entropy. 
\item therefore, it can not be identified with entropy of entanglement  which is intrinsically positive quantity.
\item this contribution does not vanish  even in two dimensions when the entropy of entanglement 
is identically zero as no physical propagating degrees of freedom are present in the system;
\item  this contribution   is gauge invariant in two dimensions and gauge dependent in four dimensional case \cite{Iellici:1996gv};
\item the technical reason for this phenomenon to happen is as follows.  One can not use the physical Coulomb gauge 
(when only physical   degrees of freedom are present in the system) as it breaks down at the origin, where $A_{\theta}$ is ill defined. Therefore, an alternative  description  in terms of   a covariant gauge  (when unphysical   degrees of freedom inevitably appear in the system) should be used instead. 
\item in this covariant description  the entropy is obtained by varying the path integral with respect to the deficit angle of the cone
as explained in~\cite{Kabat:1995eq}. Such a procedure  can (in principle) lead to a negative value for the entropy. In fact, it does come out negative~\cite{Kabat:1995eq}. 
\end{enumerate}
The main goal of the present work is to argue that the presence of this  ``weird"  term is intimately related to the  well known  property of gauge theories when a summation over all topological sectors   must  be performed for  the path integral to be  properly defined.  We explain how  all features  a)-g) from the list above can be  natural understood  within our framework when summation over topological sectors is properly taken into account. In Minkowski space the corresponding procedure is known to produce a  non-dispersive contact term with a ``wrong sign" which plays a crucial role in resolution of the so-called $U(1)_A$ problem in QCD. Exactly this non-dispersive contact term    eventually becomes the ``weird"  term with properties a)-g) when we go  from conventional Minkowski space   into curved/time-dependent background with a causal horizon.

Our consideration in this paper will be based on analysis of  the local characteristics (such as topological susceptibility, 
free energy density, etc)  computed deep inside  the horizon region. It is very different from
the computation of a global characteristic such as the total entropy   for a black hole when the  closest  vicinity  of the horizon (just outside of it) plays the crucial role in the computations.   Nevertheless, we    remain   sensitive to  the existence of  horizon because our analysis  is based on consideration of  some specific   topologically- protected quantities. Essentially, by analyzing very unusual features  a)-g)   listed above  we    learn  some important  lessons  on  behaviour   of the ground state  resulting from  merely    existence of a    causal horizon in the presence of  the gauge degrees of freedom  in the system.

The workflow is as follows. In section \ref{2d} we present our arguments for two dimensional case when all computations can be  explicitly performed. We generalize our arguments for four dimensional case in section \ref{4d}. We  argue that this term is indeed gauge dependent 
in four dimensions in abelian case as explicit computations of ref. \cite{Iellici:1996gv} suggest. However, we shall argue that this term becomes a gauge independent in non-abelian case.  
We make few comments on some  profound consequences of our findings in section \ref{consequences}, 
where we speculate that the source of the observed dark energy might be related to  the same gauge configurations which are responsible for the mismatch between black hole entropy and the entropy  of entanglement. 

\section{Topological sectors, contact term with ``wrong sign",  and all that for 2d QED
 in Rindler Space
}\label{2d}
First of all, we shall  demonstrate below the presence of a nonconventional contribution into the energy  with a ``wrong sign" in Minkowski space.
This contribution is gauge invariant, it exists even in pure  photo-dynamics when no propagating degrees of freedom are present in the system. It can not be removed by any means (such as redefinition of the energy) as it is a  real physical contribution. In particular, the anomalous Ward Identities (which emerge when the  massless fermions are added  into the system) can not be satisfied without this term.
We shall argue that this term can be treated as a contact term, and in fact is   related to the existence of  different topological sectors in this (naively trivial) two dimensional photo-dynamics. In different words, the presence of different topological sectors 
in the system, which we call the ``degeneracy" for short\footnote{Not to be confused with conventional term ``degeneracy" when two or more physically distinct states are present in the system. In the context of this paper the
     ``degeneracy" implies there existence of winding states $| n\ra$ constructed as follows: ${\cal T} | n\ra= | n+1\ra$.  In this formula the operator ${\cal T}$ is  the  large gauge transformation operator  which commutes  with the Hamiltonian $[{\cal T}, H]=0$. 
     The physical vacuum state is {\it unique}
     and constructed as a superposition of $| n\ra$ states. In path integral approach the presence of $n$ different sectors in the system is reflected  by  summation over $ { k \in \mathbb{Z}}$ in eq. (\ref{Z}, \ref{chi3},\ref{chi4}). }, is the source for this contact term which is not related to any physical propagating degrees of freedom. 

As the next step, we shall discuss the same system in the presence of the horizon in the Rindler space.
We shall argue that the contact term (which emerges itself as a result of topological features of gauge theory)  demonstrates the
  ``weird"  and  strange properties listed above in the presence of the horizon. 

In what follows it is convenient  to study     the topological susceptibility $\chi$ (rather than free energy itself)  which is related to the $\theta$ dependent portion of the free energy density\footnote{in case of infinite manifold (rather than finite size $\beta=T^{-1}$)    the free energy from  relation (\ref{chi}) becomes the conventional vacuum energy   
   as employed in study of the $U(1)_A$ problem in QCD  in \cite{witten,ven,vendiv}.} as follows 
\be
\label{chi}
 \chi (\beta,\theta=0)= - \left. \frac{\partial^2F_{\mathrm{vac}}(\beta,\theta)}{\partial \theta^2} \right|_{\theta=0},
 \ee
 where $\theta$ is the conventional  $\theta$ parameter which enters the  Lagrangian along with  topological density operator, see precise definition below. We always assume that $\theta=0$, however $ \chi (\theta=0)\neq 0$   does not   vanish,
and in fact is the main ingredient of the resolution of the $U(1)_A$ problem in QCD \cite{witten,ven,vendiv}, see also~\cite{Rosenzweig:1979ay,Nath:1979ik,Kawarabayashi:1980dp}.
 Free energy itself $F_{\mathrm{vac}}(\theta)$ can be always restored from $\chi$ as dependence on $\theta$ is known to be $F_{\mathrm{vac}}\sim \cos\theta$. 
  As we show below,     the topological susceptibility   $\chi$ (and therefore $F_{\mathrm{vac}}$),  being the local characteristics  
  of the system, nevertheless are quite sensitive to  merely existence  of horizon, even when computed far away from it.  As we shall see this sensitivity  is related to the degeneracy of the system and      topological nature of $\chi$.  
 
     \subsection{ Topological susceptibility and contact term  }\label{contact}
     The simplest (and physically attractive) choice is  Coulomb gauge when no physical propagating degrees of freedom are present in the system, and therefore the dynamics must be trivial.  It is well known, why this naive argument fails: 
     the vacuum in this system is degenerate, and one should consider an infinite superposition of of the winding states $| n\ra$ 
     as originally discussed in \cite{KS}. Such a construction in  Coulomb gauge restores the cluster and other important properties of quantum field theory.  The vacuum  in this gauge is characterized by long range forces (if  charged physical fermions  are introduced into the system). This long range force  prevents distant regions from acting  independently. We believe that precisely 
     this feature leads to the   difficulties mentioned  in \cite{Kabat:1995eq}  in computations of the entropy in physical Coulomb gauge  in two dimensions, where a covariant gauge has been used instead. 
          
     As our goal is to make a connection    with computations of ref. \cite{Kabat:1995eq}  we shall not elaborate on Coulomb gauge in the present paper any further, but rather  consider a covariant gauge to study this system. In the covariant Lorentz gauge there are no long range forces. Instead,  a new (unphysical) degrees of freedom emerge in the system, see precise definition below. 
     
 We want to study   the topological susceptibility $\chi$ in the Lorentz gauge defined as follows\footnote{Here   we use Euclidean metric where path integral computations (\ref{exact}) have been performed.}, 
\be
\label{chi1}
\chi \equiv \frac{e^2}{4\pi^2} \lim_{k\rightarrow 0} \int \dd^2x e^{ikx}\left< T E(x) E(0) \right> ,
\ee
where $Q=\frac{e}{2\pi}E$ is the topological charge density 
and 
\be
\label{k}
\int  \dd^2x ~Q(x)= \frac{e}{2\pi} \int \dd^2x ~E(x) =k
\ee
is the integer valued topological charge   in the 2d $U(1)$ gauge theory, $E(x)=\partial_1A_2-\partial_2A_1$ is the field strength. 
  The  expression for the topological susceptibility in 2d Schwinger QED model is known exactly \cite{SW}
 \be
\label{exact}
\chi_{QED}= \frac{e^2}{4\pi^2}  \int   \dd^2x \left[ \delta^2(x) - \frac{e^2}{2\pi^2} K_0(\mu |x|) \right] ,
\ee
where $\mu^2=e^2/\pi$ is the mass of the single physical state in this model, and $K_0(\mu |x|) $ is the modified Bessel function of order $0$, which 
is the Green's function of this   massive particle. 
The expression for $\chi$ for pure photo-dynamics is given by (\ref{exact}) with coupling  $e=0$ in the brackets\footnote{factor $\frac{e^2}{4\pi^2}$ in front  of (\ref{exact}) does not vanish in this limit as it is due to our definition (\ref{chi1}) rather than result of dynamics} which corresponds to the de-coupling from matter field $\psi$, i.e.
\be
\label{exact1}
\chi_{E\&M}= \frac{e^2}{4\pi^2}  \int   \dd^2x \left[ \delta^2(x)  \right] .
\ee
The crucial observation here is as follows: any physical state contributes to $\chi$ with negative sign 
\be
\label{dispersion}
\chi_{dispersive} \sim  \lim_{k\rightarrow 0} \sum_n  \frac{\la 0| \frac{e}{2\pi}   E  |n\ra \la n | \frac{e}{2\pi}   E |0\ra }{-k^2-m_n^2} <0.
\ee
In particular, the term proportional $ -K_0(\mu |x|) $ with negative sign  in eq. (\ref{exact}) is  resulted from the only physical field  of   mass $\mu$. 
However,   there is also a contact term $ \int   \dd^2x \left[ \delta^2(x)  \right]$ in eqs. (\ref{exact}), (\ref{exact1}) which contributes to the topological susceptibility $\chi$  with the {\it opposite sign}, and which can not be identified according to (\ref{dispersion}) with any contribution from any physical asymptotic state. 

This term has  fundamentally different, non-dispersive  nature. In fact it is ultimately related to different topological sectors of the theory and the degeneracy of the ground state~\cite{Zhitnitsky:2010ji} as we shortly review below.
   Without this   contribution it would be impossible to satisfy the Ward Identity (WI) because the physical propagating degrees of freedom can only contribute with sign $(-)$ to the correlation function as eq. (\ref{dispersion}) suggests, while WI requires $\chi= 0$  in the  chiral limit $m=0$.
 One can explicitly check that WI is indeed automatically satisfied\footnote{When $m\neq 0$ the WI takes the form $\chi\sim m|\la \bar{\psi}\psi\ra|$.  It  is also automatically satisfied because  $\mu^2= \frac{e^2}{ \pi} +{\cal O}(m)$, and cancellation in eq. (\ref{chi2}) is not exact resulting in behaviour    $\chi\sim m$ in complete accord with WI~\cite{Urban:2009wb}.}  only as a result of exact cancellation between conventional dispersive term with sign $(-)$ and non-dispersive term (\ref{exact1}) with sign $(+)$,   
 \be
\label{chi2}
\chi  &=& \frac{e^2}{4\pi^2}  \int   \dd^2x \left[ \delta^2(x) - \frac{e^2}{2\pi^2} K_0(\mu |x|) \right] \\  \nonumber
&=& \frac{e^2}{4\pi^2} \left[ 1- \frac{e^2}{\pi}\frac{1}{\mu^2}\right]= \frac{e^2}{4\pi^2} \left[ 1-1\right]=0.
\ee

 \subsection{The origin of the contact term-- summation   over    topological sectors.}\label{sectors}
 The goal here is to demonstrate that the contact term in exact formulae (\ref{exact}), (\ref{exact1})   is a result of the summation over different topological sectors  in the 2d pure $U(1)$ gauge theory as we now show. 
 We follow \cite{SW} and   introduce the  classical ``instanton potential"
in order to describe   the  different  topological sectors of the theory  which are classified  by integer number $k$ defined in eq. (\ref{k}). The corresponding configurations   in the Lorentz gauge on two dimensional Euclidean torus   with total area $V$ can be described as follows \cite{SW}:
 \be
 \label{instanton}
 A_{\mu}^{(k)}=-\frac{\pi k}{e V}\epsilon_{\mu\nu}x^{\nu}, ~~~ e E^{(k)}=\frac{2\pi k}{V}, 
 \ee
 such that the action of this classical configuration is
 \be
 \label{action}
 \frac{1}{2}\int d^2x E^2= \frac{2\pi^2 k^2}{e^2 V}.
 \ee
   This configuration corresponds to the   topological charge $k$ as defined by (\ref{k}).
The next step is to  compute   the topological susceptibility for the theory defined by the following partition function
\be
\label{Z}
{\cal{Z}}=\sum_{ k \in \mathbb{Z}}{\int {\cal{D}}}A {e^{-\frac{1}{2}\int d^2x E^2}}.
\ee
   All integrals in this partition function are gaussian and can be easily evaluated using the technique developed in \cite{SW}.
   The result     is determined  essentially  by the classical configurations (\ref{instanton}), (\ref{action}) as real propagating degrees of freedom are not present in the system    of pure $U(1)$ gauge  field theory in two dimensions. We are interested in computing $\chi$ defined by eq. (\ref{chi1}). In path integral approach it can be represented as follows, 
   \be\label{chi3}
\chi =\frac{e^2}{4\pi^2 \cal{Z}}\sum_{k\in \mathbb{Z}}{\int{\cal{D}}}A\int\dd^2x E(x) E(0){e^{-\frac{1}{2}\int d^2x E^2}}.  
\ee
This gaussian integral can be easily evaluated\footnote{One can check that  the contribution resulting from the quantum fluctuations about the classical background (\ref{instanton}) does not change the result (\ref{chi5}). Indeed, the corresponding extra ``quantum" contribution 
  $\frac{e^2}{4\pi^2} \cdot \int d^2x [\delta^2(x)-\frac{1}{V}]=0$ vanishes as expected.}   and  the result can be represented as follows~\cite{Zhitnitsky:2010ji}, 
   \be\label{chi4}
\chi = \frac{e^2}{4\pi^2} \cdot V\cdot \frac{  \sum_{ k \in \mathbb{Z}} \frac{4\pi^2k^2}{e^2 V^2}  \exp(-\frac{2\pi^2 k^2}{e^2 V})}{ \sum_{ k \in \mathbb{Z}}  \exp(-\frac{2\pi^2 k^2}{e^2 V})}.
\ee
 In the large volume limit $V\rightarrow \infty$ one can evaluate the sums entering (\ref{chi4}) by replacing    $ \sum_{ k \in \mathbb{Z}}\rightarrow \int d k $
 such that the leading term in eq. (\ref{chi4}) takes the form
     \be\label{chi5}
\chi= \frac{e^2}{4\pi^2} \cdot V\cdot  \frac{4\pi^2}{e^2 V^2} \cdot\frac{e^2 V}{4\pi^2}=  \frac{e^2}{4\pi^2}.
\ee
Few comments are in order. First, the obtained expression for the topological susceptibility 
(\ref{chi5}) is finite in the  limit $V\rightarrow \infty$,  coincides with the contact term from 
exact computations (\ref{exact}), (\ref{exact1}) performed for 2d Schwinger model in ref.~\cite{SW} and has ``wrong" sign in comparison with any physical contributions (\ref{dispersion}).  Second, the topological sectors with very large $k\sim \sqrt{e^2V}$ saturate the series (\ref{chi4}). 
As one can see from the computations presented above, the final result  (\ref{chi5}) is sensitive to the boundaries, infrared regularization, and many other aspects which are normally 
ignored when a  theory from the very beginning is formulated in infinite space with  conventional  assumption about  trivial behaviour at infinity. 
  Last, but not least: the contribution (\ref{chi5}) does not vanish in a trivial  model when  no any propagating degrees of freedom are present in the system!  This term is entirely determined by the behaviour at the boundary, which is conveniently  represented by  the classical topological  configurations (\ref{instanton}) describing  different topological sectors (\ref{k}), and accounts for the degeneracy of the ground state\footnote{See footnote 1 for clarification of the term ``degeneracy". In the given context the degeneracy implies the summation over $ { k \in \mathbb{Z}}$ in eq. (\ref{Z}, \ref{chi3},\ref{chi4}).}.
   We know that this term  must be present in the theory 
when the dynamical quarks are introduced into the system. Indeed,    it   plays a crucial role in this case as it saturates the WI as formula  (\ref{chi2}) shows.

 \subsection{The ghost as a tool to describe the contact term}\label{ghost}
 The goal here is to describe exactly the same contact  term (\ref{exact1}, \ref{chi5}) without explicit summation over different topological sectors, but rather, using the auxiliary ghost fields  as it has been originally done in ref. \cite{KS} (using the so-called  the Kogut-Susskind dipole). This auxiliary ghost field effectively accounts for the degeneracy of the ground state as discussed above. The computations in both refs.  ~\cite{Kabat:1995eq} and ~\cite{KS}
 are performed precisely in terms of the same auxiliary scalar field defined as  follows
 \be
 \label{A}
 A_{\mu}=\epsilon_{\mu\nu}\partial^{\nu}\Phi.
 \ee
 This formal connection   allows us to make a link 
 between expressions (\ref{exact1}, \ref{chi5}) for the contact term with ``wrong" sign computed in 
 our framework in terms of the auxiliary scalar field  as described below
  and the entropy computations performed in ref.~\cite{Kabat:1995eq} 
   featuring the ``weird properties"  e)-g) as listed in Introduction.    
 
 Our starting point   is the effective   Lagrangian describing the same two dimensional gauge system. 
However, now the theory is formulated  in covariant Lorentz gauge  in  terms of the scalar fields ~\cite{KS}. The crucial element accounting for  different topological sectors
of the underlying  theory, and corresponding degeneracy of the ground state,  does not go away    in this description. Rather,  this information is now coded  in terms of 
unphysical ghost scalar field which provides  the required ``wrong" sign for  contact  term (\ref{exact1}, \ref{chi5}).

Precise construction goes as follows.    The  effective   Lagrangian describing the low energy physics
(in Minkowski metric) is given by
~\cite{KS}:
 \be\label{lagC}
{\cal L} &=&    \frac{1}{2} \partial_\mu \hat\phi  \partial^\mu \hat\phi  +\frac{1}{2} \partial_\mu \phi_2 \partial^\mu \phi_2 - \frac{1}{2} \partial_\mu \phi_1 \partial^\mu \phi_1    \\ \nonumber
 &-& \frac{1}{2}\mu^2 \hat\phi^2   +{ m |\<\bar{q}q\>|}  \cos 2\sqrt{\pi} \left[ \hat\phi+ \phi_2 - \phi_1 \right] \, .
\ee
The fields appearing in this Lagrangian are
\be\label{names}
 \phi_1 = ~\mathrm{the~ ghost}, ~~ \phi_2 = ~\mathrm{its~ partner} \, ,
\ee
while $\hat\phi$ is the only physical massive degree of freedom.
It is important to realize that the ghost field $\phi_1$ is always paired up with $\phi_2$ in each and every gauge invariant matrix element, as explained in ~\cite{KS}.  The condition that enforces this statement is the Gupta-Bleuler-like condition on the physical Hilbert space ${\cal H}_{\mathrm{phys}}$ which reads like
\be\label{gb}
(\phi_2 - \phi_1)^{(+)} \left|{\cal H}_{\mathrm{phys}}\right> = 0 \, ,
\ee
where the $(+)$ stands for the positive frequency Fourier components of the quantized fields.  
One can easily understand the origin for a wrong sign for the kinetic term for $\phi_1$ field.  
 It occurs  as a result of $ \Box^2$ operator when the Maxwell term $E^2\sim  \Box^2$ is expressed in terms of the scalar field (\ref{A}).  As usual, the presence of 4-th order operator  is a signal that the ghost is present in the system.   Indeed, the relevant operator $ \left[  \Box \Box + \mu^2\Box \right] $ which emerges for this system can be represented as the combination of the ghost $\phi_1$ 
and a massive physical $\hat\phi $ using the standard trick by 
  writing  the inverse operator as follows
\be
\label{inverse}
\frac{1}{\Box \Box +\mu^2\Box  }= \frac{1}{\mu^2}\left(\frac{1}{ -\Box - \mu^2  } -  \frac{1}{ - \Box  } \right) \, .
\ee
 This is a simplified explanation on how sign $(-)$ emerges in Lagrangian (\ref{lagC}) describing auxiliary $\phi_1$ field, see original paper\cite{KS} for details. 
 
 The   contact term     in this framework is precisely represented 
by the ghost contribution\cite{Zhitnitsky:2010ji,Urban:2009wb} replacing the standard procedure 
of summation over different topological sectors as discussed above \ref{sectors}.
 Indeed,   the topological density $Q=\frac{e}{2\pi}E$ in 2d QED is given by $\frac{e}{2\pi} E= (\frac{e}{2\pi}){\frac{\sqrt{\pi}}{e}}\left(  \Box\hat\phi -  \Box\phi_1 \right)$ ~\cite{KS}.
 The relevant correlation function in coordinate space which enters the expression for the topological susceptibility (\ref{chi1}) can be explicitly computed using the ghost as follows 
 \be\label{chi6}
\chi(x) &\equiv& \left< T   \frac{e}{2\pi} E(x) , \frac{e}{2\pi} E(0) \right>   \nn \\
&=&\left( \frac{e}{2\pi} \right)^2\frac{\pi}{e^2} \int \frac{\dd^2p}{\left(2\pi\right)^2} p^4 e^{-i p x} \left[ - \frac{1}{p^2+\mu^2} + \frac{1}{p^2} \right] \nn \\
  & =& \left(\frac{e}{2\pi} \right)^2\left[ \delta^2(x) - \frac{e^2}{2\pi^2} K_0(\mu |x|)\right]    
  \ee
where we used the known expressions for the  Green's functions (the physical massive field $\hat\phi $ as well as the ghost  $\phi_1 $ field) determined by Lagrangian (\ref{lagC}) and switched back to Euclidean metric for   comparison with previous  results from sections \ref{contact},~\ref{sectors}.

  The obtained expression   precisely reproduces exact   result (\ref{exact}) as claimed.  In the limit $e\rightarrow 0$ when  matter fields decouple from gauge degrees of freedom  we reproduce the contact term  (\ref{exact1}, \ref{chi5}) which was previously derived as a result of summation over  different topological sectors of the theory. The non-dispersive contribution manifests itself in this description 
   in terms of 
unphysical ghost scalar field which provides  the required ``wrong" sign for  contact  term.

At the same time, this unphysical ghost scalar field  does not violate unitarity or any other important 
properties of the theory as  consequence  of Gupta-Bleuler-like condition on the physical Hilbert space  (\ref{gb}).  
Indeed, 
while the   ghost's number density operator, $\mathrm{N} $  may look dangerous due to the sign $(-)$ in the commutation relations
\be
\label{H}
\mathrm{N} &=&\sum_k  \left(b_k^{\dagger}b_k- a_k^{\dagger} a_k\right)  \\
 \left[b_k, b_{k'}^{\dagger}\right]&=&\delta_{kk'} ,  \left[a_k, a_{k'}^{\dagger}\right]=-\delta_{kk'}    \nn
\ee
one can in fact check that  the expectation value for any physical state   vanishes as a result of the subsidiary condition~\cite{KS,Zhitnitsky:2010ji}:
\be
\label{H=0}
\< {\cal H}_{\mathrm{phys}}| \mathrm{N} |{\cal H}_{\mathrm{phys}}\>=0 ~, 
(a_k-b_k) \left|{\cal H}_{\mathrm{phys}}\right> = 0.~~ 
\ee
 This vanishing result (\ref{H=0})  obviously implies that no  entropy  may  be produced in Minkowski space. In different words, the   fluctuations   of unphysical fields described by operator (\ref{H}) do not lead to  any physical consequences (except for merely existence of the  contact term (\ref{exact1}) as already discussed).
 
We shall see in next subsection  how this  simple picture  drastically changes when we 
 consider the very same system but in  presence of the horizon. We shall argue that the number density $N$ of ``fictitious 
 particles"  with wrong commutation relations starts to fluctuate in the presence of the horizon, in contrast with eq. (\ref{H=0}). Therefore, 
we formulate a conjecture   that precisely  these fluctuations are responsible for  term with a ``wrong sign" in entropy computations 
 ~\cite{Kabat:1995eq,Iellici:1996gv}. The corresponding contribution, as we already mentioned, is not related to any physical propagating degrees of freedom but rather, is due to presence of topological sectors in gauge theories
 (and   the degeneracy of the ground state as its consequence, see footnote 1 for clarification on terminology)  which eventually lead to a non-dispersive contribution in topological susceptibility.  
  To simplify things in what follows we consider a simple  Rindler space when the   Bogolubov's coefficients are exactly known. However, we argue that a generic case  (when  horizon is present in the system) leads to very similar conclusion.  
 \subsection{Rindler space}\label{Rindler}
 The total entropy with ``weird" properties listed in Introduction was computed a while ago ~\cite{Kabat:1995eq}, and there is no reason 
 to review these results in the present paper. These original  results  have been reproduced in \cite{Iellici:1996gv} by using another technique. Furthermore, in the same paper  \cite{Iellici:1996gv}  it has been demonstrated that   in two dimensional case the final result is gauge invariant, and therefore, it obviously represents a physically observable quantity. As we mentioned earlier, we are not interested in computing the global characteristics such as total entropy.   Rather, we are interested in computing some local   properties, such as topological susceptibility, or $\theta-$ dependent portion of the energy density (\ref{chi}) in the presence of the horizon. However, we shall argue below, the source for ``weird" features in both 
 cases is the same, and, in fact,  related to fundamental properties of gauge theories as discussed   in section \ref{sectors}.
 
 As we explained  above, the presence of different topological sectors in gauge theory 
  (and   the degeneracy of the ground state as its consequence) leads to contact term (\ref{exact1})
 even when no physical propagating degrees of freedom are present in the system. In physical Coulomb gauge this term manifests itself as the presence of  long range force which prevents distant regions from acting independently. The same feature 
but in  covariant Lorentz gauge is expressed in terms of  new
(unphysical) degrees of freedom (\ref{names}) which emerge in the system and effectively reproduce the contact term as explicit computations show (\ref{chi6}). While these unphysical degrees of freedom do fluctuate, these fluctuations do not lead to any physical observable expectation values in Minkowski space (\ref{H=0}) as a result of cancellation between two unphysical fields  similar to conventional Gupta-Bleuler condition in QED when two unphysical photon's polarizations cancel each other. 
 We want to see how this conclusion changes when a horizon is present in the system. 
 
 One can repeat the construction of previous section \ref{ghost} to describe (unphysical) degrees of freedom but in Rindler space~\cite{Zhitnitsky:2010ji}. 
  A Rindler observer in (R,L) wedge will measure the number density of unphysical states  using  density operator $\mathrm{N}^{(R,L)}$ which is given by
\be
\label{H-R}
\mathrm{N}^{(R,L)}=\sum_k  \left(b_k^{(R,L)\dagger}b_k^{(R,L)}- a_k^{(R,L)\dagger}a_k^{(R,L)}\right) \, .
\ee
The subsidiary condition~(\ref{gb}) defines the physical subspace for accelerating Rindler observer 
\be\label{gb-R}
\left(a_k^{(R,L)}-b_k^{(R,L)}\right) \left|{\cal H}_{\mathrm{phys}}^{(R,L)}\right> = 0 \, ,  
\ee
such that the exact cancellation between $\phi_1$ and $\phi_2$ fields holds  for any physical state defined by eq. (\ref{gb-R}), i.e.
\beq
  \left<{\cal H}_{\mathrm{phys}}^{(R,L)}|\mathrm{N}^{(R,L)}  |{\cal H}_{\mathrm{phys}}^{(R,L)}\right> = 0 
  \eeq
as it should.
However, if the system is prepared as the Minkowski vacuum state $ |0\> $  then  a Rindler observer using the same operator for $\mathrm{N}^{(R,L)}$
(\ref{H-R}) will observe the following number density in mode $k$, 
\be
\label{RR}
\< 0 | \mathrm{N}^{(R,L)} |0\>&=&\< 0 | \left(b_k^{(R,L)\dagger}b_k^{(R,L)}- a_k^{(R,L)\dagger}a_k^{(R,L)}\right)  |0\>
 \nonumber  \\
 &=& \frac{2   e^{-\pi\omega/a}}{{(e^{\pi\omega/a}-e^{-\pi\omega/a})}}= \frac{2  }{(e^{2\pi\omega/a}-1)},
\ee
where we used known Bogolubov's coefficients mixing the positive and negative  frequency  modes for   operators $b_k^{(R,L)}, a_k^{(R,L)}$ describing unphysical fluctuations~\cite{Zhitnitsky:2010ji}.
 
   One can explicitly see why the cancellation (\ref{H=0}) of unphysical degrees of freedom in Minkowski space fail  to hold 
 for the accelerating  Rindler observer (\ref{RR}).  
 The technical reason for this effect to occur  is the property of Bogolubov's coefficients which mix the positive and negative frequencies modes. The corresponding  mixture can not be avoided because  the projections to 
     positive -frequency modes 
  with respect to Minkowski time $t$  and  positive -frequency modes    with respect to the Rindler observer's proper  time $\eta$  are  not equivalent. 
  The exact cancellation of unphysical degrees of freedom which is maintained in Minkowski space can not hold in the Rindler space because  it would be not possible to separate positive frequency modes from negative frequency ones in the entire spacetime, in contrast with what happens in Minkowski space where the vector $\partial/\partial t$ is a constant  Killing vector, orthogonal to the $t=\mathrm{const}$ hypersurface.  The Minkowski separation is maintained throughout the whole  space as a consequence of Poincar\'e invariance. It is in a drastic contrast with  the accelerating  Rindler  space~\cite{Zhitnitsky:2010ji}.  
  
  The nature of the effect   is    the same as  the  conventional Unruh effect\cite{Unruh:1976db} when the  Minkowski  vacuum $ \left| 0 \right>$ is restricted to the Rindler wedge  with no   access to the entire space time.  An appropriate description in this case, as is known,    should be  formulated (for $R$ observer) in terms of the density matrix by ``tracing out" over the degrees of freedom associated with $L$-region. In this case the  Minkowski  vacuum $ \left| 0 \right>$ is obviously not a pure state but a mixed state with a horizon separating two wages, which is the source of the entropy. 
 In contrast with Unruh effect\cite{Unruh:1976db}, however,  one can not speak about real radiation of real particles as 
 the ghost $\phi_1$ and its partner $\phi_2$ are not the asymptotic states 
  and the corresponding positive frequency 
    Wightman Green function describing the dynamics of these fields vanishes ~\cite{Zhitnitsky:2010ji}. In different words, 
    these auxiliary fields contribute to the non-dispersive portion of the correlation function  in eqs. (\ref{exact},\ref{exact1},\ref{chi2}), but not to conventional dispersive part  
    which is unambiguously determined by the absorptive function as conventional dispersion relation dictates.

  Few more comments on (\ref{RR}) are in order.  The effect is obviously sensitive to the presence of the horizon, and, therefore is  infrared (IR) in nature. 
 The IR nature of the effect was anticipated from the very beginning as formulation of the problem  in terms of auxiliary fields (\ref{names})  is simply a convenient way to deal with different topological sectors of the gauge theory in covariant gauge 
  (and   the degeneracy of the ground state as their consequence)
 instead of dealing with the long range forces in the unitary Coulomb gauge  as discussed in sections \ref{sectors} and  \ref{ghost}. 
 Also:  the contribution  of higher frequency modes are exponentially suppressed $\sim \exp(-\omega/a)$ as expected. 
 The interpretation of eq. (\ref{RR}) in terms of particles is very problematic (as usual for such kind of problems)  as typical frequencies  when the effect (\ref{RR}) is not exponentially small, are of order $\omega\sim a$, and notion of ``particle" for such $\omega$  is not well defined.

 We do not attempt  to reproduce known  results on entropy from ref.~\cite{Kabat:1995eq}  based on non-vanishing expectation value  for number density operator (\ref{RR}).
 First of all,  it is not obvious what would be   the physical meaning of such a computation based on expectation value (\ref{RR}) for the  operator which satisfies
 ``wrong" commutation relation (\ref{H}). Furthermore, it is not obvious  how to interpret  $  \mathrm{N}   $ particles from  eq.(\ref{RR}) when entire notion of ÒparticlesÓ is not even defined for relevant parameters.   
 Indeed, as we argued above  the effect is large  $\< 0 | \mathrm{N}  |0\> \sim 1$ only for very large wave length $\lambda\geq a^{-1}$ which is the size of the horizon scale.
 
 Our goal here is in fact  quite different. We want to argue that the source for the ``wrong sign" in   entropy computations ~\cite{Kabat:1995eq} (featuring  
the ``weird properties"    as listed in Introduction) and the source for the ``wrong sign" for the contact non-dispersive term (discussed in present paper)  are in fact  have the same origin. 
 In addition to the arguments presented above, we note that the technical computations of the entropy performed in 
 ~\cite{Kabat:1995eq}   are actually based precisely on the  same representation  for $A_{\mu}$  field (\ref{A})  describing fluctuations of unphysical auxiliary degrees of freedom. This representation 
   for $A_{\mu}$  field in our formalism       eventually leads 
   to the expression for the contact non-dispersive contribution  $\sim \delta^2(x)$ with ``wrong" sign  (\ref{chi6}) 
   and non vanishing number density (\ref{RR}) while in ref.\cite{Kabat:1995eq}  the very same representation  for $A_{\mu}$  field (\ref{A}) leads to the ``wrong" sign for the entropy. Furthermore, the contact term (\ref{exact1}) can be represented as a surface term,
   \be
   \label{div}
 \chi_{E\&M}\sim  \int   \dd^2x \left[ \delta^2(x)  \right] =   \int   \dd^2x~
 \partial_{\mu}\left(\frac{x^{\mu}}{2\pi x^2}\right),
    \ee
    analogous to the ``weird" contribution in the  entropy computations~\cite{Kabat:1995eq,Iellici:1996gv}.
  It is important to realize that  the contact term (\ref{exact1}, \ref{chi5}, \ref{div}) is a result of summation over all 
   topological sectors with inclusion of  all  quantum fluctuations which account for the degeneracy of the ground state as discussed in section \ref{sectors}. At the same time,  quite miraculously,  the final result  (\ref{exact1}, \ref{chi5}, \ref{div}) can be interpreted  as a surface integral of a single  classical configuration of a  pure gauge field  $A_{\mu}^{cl}\sim \partial_{\mu}\phi^{cl}$ defined on a distance surface $S_1$ and characterized by unit winding number
   \be
   \label{winding}
   \chi_{E\&M}\sim
  \oint_{S_1}  \frac{A_{\mu}^{cl} dl^{\mu}}{2\pi}=\oint_{S_1}  \frac{rA_{\theta}^{cl} d\theta}{2\pi}
  =\int  \frac{d\theta}{2\pi}\frac{\partial\phi^{cl}}{\partial\theta}.~~
   \ee
   These   observations strongly suggest that the term with a ``wrong" sign in the expression for  entropy derived in ~\cite{Kabat:1995eq}  has exactly the same {\it origin} 
     as the ``wrong" sign  for the contact term  (\ref{exact1})  as 
     in both cases the relevant physics  is determined by the   surface integrals, not related to any physical propagating    dynamical degrees of freedom. Furthermore, 
   in both cases  the sign of the effect is opposite to what one should expect from physical degrees of freedom, and, finally, in both cases the starting point (formal representation for  $A_{\mu}$  field (\ref{A})) is the same. 
     
  $\bullet$ Therefore, we {\it conjecture} that:  the surface term with a ``wrong sign" in  entropy 
computations~\cite{Kabat:1995eq,Iellici:1996gv}  and the  ``wrong sign" in topological susceptibility (\ref{chi1},\ref{exact1},\ref{chi5})   are both originated form the same physics,  and both related to the same (topologically   nontrivial) gauge configurations, and must be present (or absent) in both computations simultaneously. In both cases a ``wrong sign" emerges  due to unphysical degrees of freedom fluctuating in far infrared (IR) region. 
   The technical treatments of these terms  in our framework and  in  ref.~\cite{Kabat:1995eq} of course is very different:   we use conventional Hamiltonian approach supplemented by the condition  (\ref{gb}) while in ref. ~\cite{Kabat:1995eq}  the Rindler Hamiltonian is ill defined on the cone, and computations are performed using some alternative methods.     
  Nevertheless, in our framework, we interpret the fluctuations (\ref{RR}) of  ``fictitious particles" with ``wrong" commutation relations (\ref{H}) as a different manifestation of the same physics which led to a Òwrong signÓ in entropy computations ~\cite{Kabat:1995eq,Iellici:1996gv}. 
   An additional argument 
   supporting our {\it conjecture}    will be  presented in the next section where we show that these very different quantities  nevertheless behave very similarly when the system is generalized from two to four dimensions, and therefore, they must be originated from the same physics.
     
     Our final  comment here is  this: 
     the IR physics   penetrates into the physical gauge invariant correlation function (\ref{chi1}) not due to the 
  massless degrees of freedom in the physical spectrum (there are none in fact), but rather, as a result of   degeneracy of the ground state and summation  over all topological sectors in gauge theory as discussed  in sections \ref{contact}, \ref{sectors}.  The ghost (\ref{names}) in this framework is simply a convenient tool to account for this far IR physics as it effectively accounts for the  non-dispersive contact term with a ``wrong sign"  (\ref{chi6}). It fluctuates in the presence of the horizon (\ref{RR}), and responsible for a ``wrong" sign in entropy computations, according to our conjecture.  However, it remains unphysical  auxiliary  field  as it  does not belong to the physical Hilbert space
  (and it never becomes  an asymptotic state capable to propagate to infinity)~\cite{Zhitnitsky:2010ji}. 
  It is interesting to notice that there are other known examples when the degeneracy of the ground state in   the presence of the horizon leads to mismatch between black hole entropy and entropy of entanglement, see Appendix \ref{subtleties} for   references and details.

\section{Generalization to 4-D case}\label{4d}

The goal of this section is twofold. First, in next subsection \ref{QED} we   make   few comments on generalization of two dimensional results discussed above   to four dimensional QED. In this case the corresponding calculations of the entropy are known~\cite{Kabat:1995eq,Iellici:1996gv}. Analysis of these results  further support  our conjecture on common nature of the surface term with a ``wrong sign" in  entropy 
computations ~\cite{Kabat:1995eq,Iellici:1996gv} and ``wrong sign" in topological susceptibility as the behaviour of the system follows precisely the pattern dictated  by the {\it conjecture}. Secondly, in section \ref{QCD} we   discuss   four dimensional non-abelian gauge theories when corresponding computations of the entropy  are not yet known. Nevertheless, based on our conjecture on common origin of these two different phenomena, 
we    predict  a possible outcome if the corresponding computations are performed.  

\subsection{Four dimensional abelian  QED}\label{QED}

We start by reviewing the basic results of refs.~\cite{Kabat:1995eq,Iellici:1996gv}
on entropy computations in four dimensional case. In   original paper ~\cite{Kabat:1995eq} the gauge invariance of the ``surface term" has not been tested. This question has been specifically discussed  in followup paper \cite{Iellici:1996gv} where it has been demonstrated that in two dimensions
the result is indeed  gauge invariant and    coincides with the original expression found in ref.~\cite{Kabat:1995eq}. 
However, a similar analysis     in four dimensions  turned out to be   much more subtle, see details in~\cite{Iellici:1996gv}.  
In particular, it has been found that this term is  gauge dependent in four dimensional abelian case, and therefore,   it was discarded~\cite{Iellici:1996gv}. 

How one can understand such puzzling  behaviour 
of the system when one jumps from two to four dimensions? 
If one accepts our {\it conjecture} formulated above then this puzzle 
  has a very natural explanation. Indeed, the photon field in two dimensions has nontrivial topological properties formally expressed by the first homotopy group $\pi_1[U(1)]\sim \cal{Z}$. It implies the degeneracy of the ground state when each topological sector  $|n\ra$ is classified by integer number. Precisely this feature
  leads  to non vanishing topological susceptibility with a ``wrong" sign in two dimensions (\ref{exact1}). The same degeneracy   leads to nontrivial instanton solutions (\ref{instanton}) interpolating between different topological sectors which saturate the  topological susceptibility (\ref{chi5}) with ``wrong sign".
  
   In contrast to two dimensional case, in four dimensions one should not expect any contact term with a ``wrong sign" similar to (\ref{exact1}) as the third homotopy group is trivial, $\pi_3(U(1))\sim 1$, there is no degeneracy of the ground state as  there is only a single  trivial vacuum state.
   Therefore, one should not expect any non trivial surface terms  in entropy computations in four dimensions.  This expectation
   based on our conjecture  is supported by explicit computations ~\cite{Iellici:1996gv}  where it was 
   shown that  in four dimensional QED the  surface  term is   gauge dependent and must be consistently discarded.
   In fact, we consider these arguments as further support for our {\it conjecture} formulated above  as the behaviour of the system  follows  precisely the pattern dictated  by this conjecture\footnote{The argument presented above is based on observation  that $4D$  space time (where computations ~\cite{Iellici:1996gv}  have been performed) has trivial topological properties.   One can consider, instead,  less trivial case when $4D$ space time is represented, for example,  by a torus, in which case 
   the relevant homotopy group could  be  non- trivial, $\pi_1(U(1))\sim \cal{Z}$,  and  contact term with a ``wrong sign" in entropy computations may occur. In principle, this is a testable proposal. Technically, though,  it could be  quite a challenging problem.}.

  Our final comment is on interpretation of the surface term with a ``wrong" sign given in conclusion of ref. \cite{Iellici:1996gv}, where it has been suggested   that, quote, ``effective low-energy string theory which does not coincide with the ordinary QFT" in principle may produce some surface term with a ``wrong" sign.
  We want to comment here, that in fact, very ordinary QFT may produce such kind of terms, which however, are non-dispersive in  nature,  and not related to any physical propagating degrees of freedom as explained in previous section \ref{sectors} for two dimensional case. As we argue in next section, such a behaviour is not a specific feature of two dimensional physics, but in fact very generic property in four dimensions as well. However, these nontrivial properties  emerge in four dimensions only for non abelian gauge fields when the third homotopy group is non-trivial, $\pi_3[SU(N)]\sim \cal{Z}$,  the  ground state is degenerate and   each topological sector  $|n\ra$ is classified by integer number similar to two dimensional case considered in section \ref{2d}.  The contact term with a ``wrong sign"  similar to eq. (\ref{exact1})  is expected to emerge in this case
  as a result  of nontrivial  topological features of  four dimensional non-abelian gauge theories.  

\subsection {Four dimensional non-abelian  QCD}\label{QCD}

  The goal of this section is to argue that all key elements from  previous section \ref{2d}
  are also present  in four dimensional QCD. In fact, the presence of the contact term with ``wrong sign" in topological susceptibility in QCD   is a crucial  element  of  resolution of the so-called $U(1)_A$  problem~ \cite{witten,ven}. The difference with two dimensional case is that in strongly coupled QCD we can not perform exact analytical computations similar to  (\ref{exact1},\ref{chi5}). However,  one can use an effective  description in terms of auxiliary   ghost  field
  \cite{ven} to compute the  non-dispersive contribution to topological susceptibility  with ``wrong sign".  This computation which employes the Veneziano ghost\footnote{not to be confused with conventional Fadeev Popov ghosts.} is direct analog  of derivation of eq.  (\ref{chi6})  when the Kogut Susskind ghost was used.  
     Essentially, our goal here is to point out  that  the relevant features in 2d QED (discussed in section \ref{2d} when all computations can be explicitly   performed) and in 4d QCD 
  (when the final word   is expected to come from the lattice numerical computations) are almost identical. To further support these similarities
 we present   some QCD lattice numerical results   explicitly measuring   the  term with a ``wrong sign" in topological susceptibility similar to eqs. (\ref{exact},\ref{chi6}). Based on these observations, our {\it conjecture} 
 essentially implies that  the entropy computations  in  four dimensional non-abelian gauge theories 
 must reveal a  contribution  with ``wrong sign" as the crucial element, the degeneracy of the ground state,
 is present in the system. Moreover, it must be  gauge invariant (and therefore, physical)    in contrast with 4D QED computations where it was shown to be  gauge variant~\cite{Iellici:1996gv}, and therefore, was discarded.

Our starting remark  is that  the expression for  topological density operator
 \be
 \label{q}
q= \partial_{\mu}K^{\mu}=\frac{g^2}{64\pi^2} \epsilon_{\mu\nu\rho\sigma} G^{a\mu\nu} G^{a\rho\sigma} =  \Box \Phi
\ee
being  represented in terms of auxiliary scalar field $\Phi$ has exactly   the same form as 
in 2d Schwinger model, 
see section \ref{ghost}. The $\Phi$ field in   formula (\ref{q}) is defined as 
$K_{\mu}\equiv \partial_{\mu} \Phi$ and is the direct analog  of representation  (\ref{A}) for 2d model.
Our next remark is that four-derivative  operator $ \int \!\dd^4x~ q^2\sim  \int \!\dd^4x ~(\Box\Phi)^2$ is expected to be induced in effective low energy Lagrangian as argued  by Veneziano \cite{ven,vendiv} in his resolution of
the $U(1)_A$ problem. 
\exclude{As is known this operator plays a key role in   the $U(1)_A$ problem, e.g. it saturates the WI and eventually generates the physical mass of the $\eta'$ field.  }
As a result of generating of  $ q^2$ operator the relevant structure which emerges in the effective Lagrangian and describing  this system is identical to  
  2d QED case, i.e. it has precisely the same structure $ \sim \Phi\left[  \Box \Box + m_{\eta'}^2\Box \right] \Phi$. The corresponding path integral  $\int{\cal D} {\cal}\Phi$ can be treated  exactly in the same way as it was treated  in 2d QED, i.e. it can be represented as the combination of the ghost $\phi_1$ 
and a massive physical $\hat\phi $ field using the same   trick by 
  writing  the inverse operator as follows
  \be
\label{inverse1}
\frac{1}{\Box \Box +m_{\eta'}^2\Box  }= \frac{1}{m_{\eta'}^2}\left(\frac{1}{ -\Box - m_{\eta'}^2  } -  \frac{1}{ - \Box  } \right) \, ,
\ee
 in complete analogy with 2d case, see eq. (\ref{inverse}).
In fact, one can show that the relevant part of the low energy QCD Lagrangian in large $N_c$ limit 
in the form suggested by Veneziano \cite{ven,vendiv} is   \emph{identical} to that proposed by Kogut and Susskind    for the 2d Schwinger model ~(\ref{lagC}), where one should replace $2\sqrt{\pi} \rightarrow f_{\eta'}^{-1}$ 
and $\mu\rightarrow m_{\eta'} $ such that the scalar fields $\phi_1, \phi_2,\hat\phi$  have appropriate (for 4D case) canonical dimension one, see \cite{dyn} for  details. 
This formal similarity leads to almost identical 
 computations   (in terms of the ghost)  of  the topological susceptibility in  4d QCD  and  in 2d QED.
 Indeed, by repeating all our previous steps leading to  eq. (\ref{chi6}) with known Green's functions which follow from (\ref{lagC}) and with known expression for topological density operator
 $q(x) \sim ( \Box\hat\phi -  \Box\phi_1)$ one arrives to the following expression for topological 
 susceptibility\footnote{
 Of course $\chi=0$ to any order in perturbation theory because $q(x)$ is a total divergence $q = \partial_\mu K^\mu $.  However, as we learnt from~\cite{ven,witten}, $\chi\neq 0$ due to the non-perturbative infrared physics.
  One can interpret   field $K^\mu $ as a unique collective mode  of the original gluon fields.  It describes the dynamics of the degenerate states $|n\ra$ representing the topologically nontrivial sectors of the ground state,   it  leads to a pole in  unphysical subspace  in the infrared,  and finally, it saturates the contact term  with a ``wrong sign" in topological susceptibility (\ref{top1}).}
 in  4d QCD in the chiral limit $m=0$,
   \be
\label{top1}
\chi_{QCD} &\equiv& \int \!\dd^4x \la T\{q(x), q(0)\}\ra \\ \nonumber
&=& \frac{f_{\eta'}^2 m_{\eta'}^2}{4} \cdot \int d^4x\left[ \delta^4 (x)- m_{\eta'}^2 D^c (m_{\eta'}x)\right], 
\ee
where $D^c (m_{\eta'}x)$ is the Green's function of a free massive particle with standard normalization $\int d^4x m_{\eta'}^2 D^c (m_{\eta'}x)=1$. In this expression the $\delta^4(x)$ represents the ghost contribution   while  the term proportional  to $ D^c (m_{\eta'}x)$ represents the physical $\eta'$ contribution, see  \cite{dyn, 4d} for details.  
The ghost's contribution can be also thought as the Witten's contact term ~\cite{witten} with ``wrong sign" which is not related to any propagating degrees of freedom. The topological susceptibility $\chi_{QCD}(m= 0)=0 $ vanishes in the chiral limit  as a result of exact cancellation of two terms entering (\ref{top1}) in complete accordance with WI.   When $m\neq 0$ the cancellation is not complete and $\chi_{QCD}\simeq m \<\bar{q}q\>$ as it should.
\begin{figure}[t]
\begin{center}
 \includegraphics[width = 0.4\textwidth]{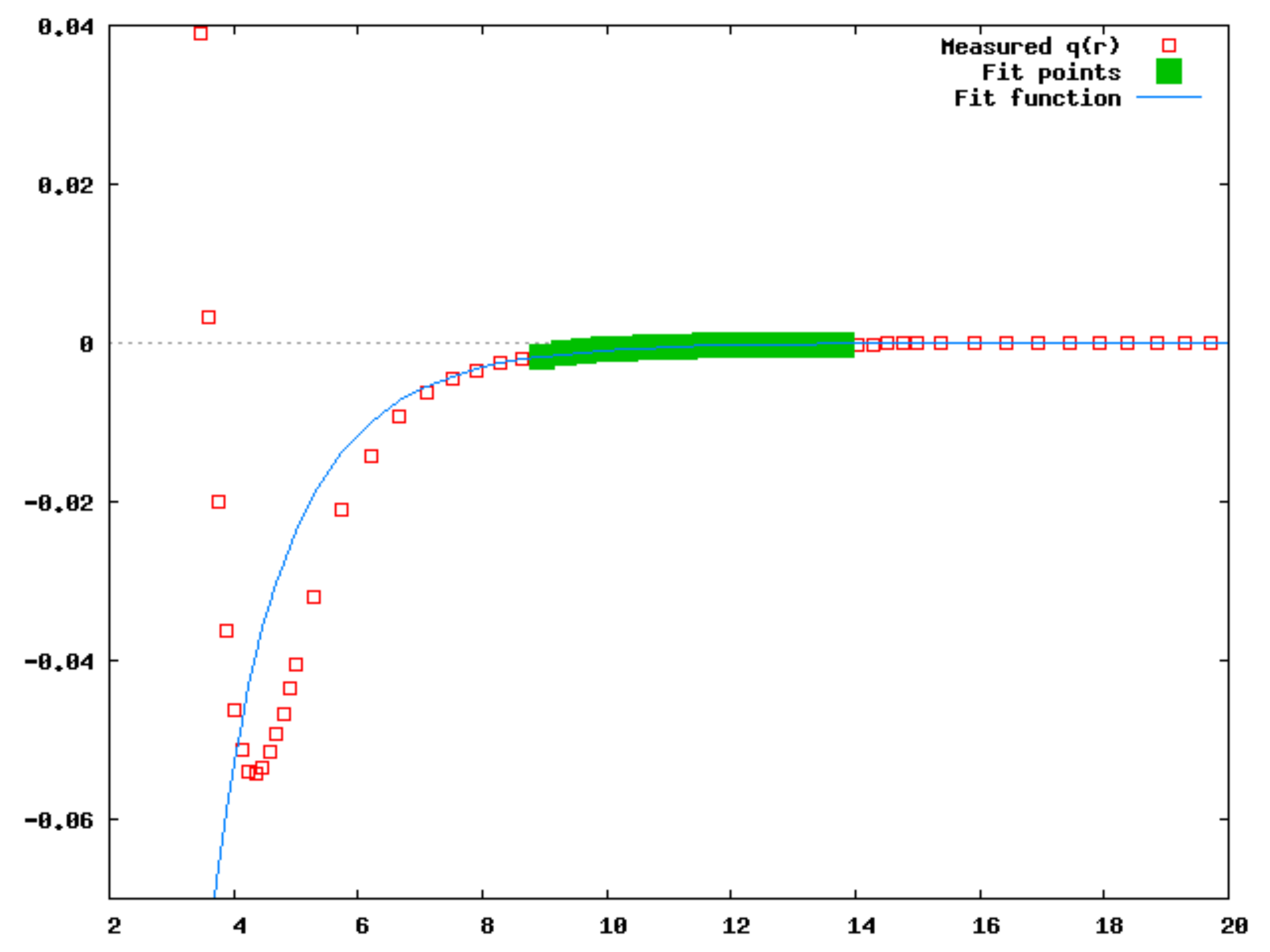}
 \caption{\label{chi-lattice}
The density of the topological susceptibility $\chi(r)\sim \< q(r), q(0)\>$ as function of separation $r$ 
such that $\chi\equiv \int d r \chi(r)$, adapted  from~\cite{lattice}. Plot explicitly shows the presence 
of the contact term with the ``wrong sign" (narrow peak around $r\simeq 0$) represented by the Veneziano ghost in our framework.  
 }
\end{center}
\end{figure}

Similar to  eq. (\ref{div}) for two dimensional system,
   the non-dispersive term with  ``wrong sign" in topological susceptibility   (\ref{top1}) in four dimensional QCD can be also represented as a surface integral 
       \be
   \label{div1}
 \chi \sim  \int   \dd^4x \left[ \delta^4(x)  \right] =   \int   \dd^4x~
 \partial_{\mu}\left(\frac{x^{\mu}}{2\pi^2 x^4}\right).
    \ee
 
In case of 2d QED we could compare our  ghost's based computations  (\ref{chi6})      with exact results (\ref{exact},\ref{exact1}) and with explicit summation over different topological sectors in pure $E\&M$ 
when no propagating degrees  of freedom are present  in the system (\ref{chi5}). 
We do not have such a  luxury in case of 4d QCD. Nevertheless, we can compare the ghost's based computations 
in 4d QCD given by eq.(\ref{top1}) with the lattice results, see e.g. 
\cite{lattice}.    We reproduce Fig.\ref{chi-lattice} from ref.\cite{lattice}  to  illustrate few    elements which are crucial for this work and which are explicitly present on the plot. First of all, there is a  narrow peak around $r\simeq 0$ with a ``wrong sign". Second,  one can observe  a smooth behaviour in  extended region  of  $r\sim \text{fm}$  with the opposite sign. Both these elements  are present
in the lattice computations as one can see from Fig.\ref{chi-lattice}. The same important elements are also present 
in our ghost's based computations given by eq. (\ref{top1}).    In different words, the QCD ghost does model 
 the crucial  property of the  topological susceptibility related to summation over topological classes in gauge theories. This feature  can not be accommodated by any physical asymptotic states as it is related to  non-dispersive contribution  in the  topological susceptibility as explained above in section \ref{sectors}, and elaborated further  in  Appendix \ref{subtleties} where this feature is explained as a result of differences in definition between the Dyson's T-product and Wick's T- product.

 Our next step is to describe the behaviour of the same system (more precisely, the behaviour of the non-dispersive term in eq. (\ref{top1}, \ref{div1}) proportional to the $  \delta^4 (x)$ function)  in Rindler space in the presence of the horizon. 
 We consider a simple case when the acceleration is sufficiently large $\Lambda_{QCD}^4\gg a^4\gg m |\<\bar{q}q\>|$
 such that   interaction  term  in  (\ref{lagC}) can be neglected and the   Bogolubov's coefficients are exactly known. 
  In this limit one can repeat all previous steps  to arrive to the same   Planck spectrum (\ref{RR}) for number density fluctuations  of 
  ``fictitious  particles"  with ``wrong" commutation relations \cite{Zhitnitsky:2010ji,Zhitnitsky:2010zx}.
  This formula (up to some irrelevant  numerical coefficient) has been reproduced in ref.\cite{ohta}
  using a different technique. 
   We interpret these fluctuations precisely in the same way as we did
  in section \ref{Rindler} in  2d  case  when  we interpreted these fluctuations of
  ``fictitious  particles"  with ``wrong" commutation relations as a different manifestation of the same physics which led  to a ``wrong sign" in entropy computations 
 ~\cite{Kabat:1995eq,Iellici:1996gv}. As we emphasized before
   the corresponding contribution  is not related to any physical propagating degrees of freedom but rather, is due to topological sectors in gauge theories which eventually lead to a non-dispersive contribution in topological susceptibility.  

   Analysis of  2d case led us to the conjecture formulated at the end of section \ref{2d} that both phenomena (``wrong sign" in entropy computations and ``wrong sign" in topological susceptibility) are originated from the same  physics determined by surface  dynamics  of the ``fictitious  particles".  In  this section we demonstrated that all relevant features   are also present in 4d non-abelian QCD. 
    Therefore, based on series of arguments presented above, it is naturally to assume that  there will be a mismatch between black hole entropy and the entropy of entanglement  
     in 4d QCD with  the same ``weird" features
  as listed in Introduction. However, in contrast  with 4d QED we expect that the surface term with ``wrong sign" in entropy will be a gauge invariant quantity similar to 2d QED  case discussed in section \ref{2d}.
  This is essentially a prediction which follows from  the {\it conjecture}. As we already mentioned there are other known examples when the degeneracy of the ground state in   the presence of the horizon leads to mismatch between black hole entropy and entropy of entanglement, see Appendix \ref{subtleties} for the details and references. One could argue that the   dynamics  of ``fictitious  particles"  
on a surface should be governed by the corresponding Chern-Simons action. We leave this subject for a future study\cite{arz}.


\section{Contact interaction and profound consequences for expanding  universe}\label{consequences}
This portion of the paper  is much more speculative in nature than the previous sections. However, these speculations may have some profound consequences on our understanding of expanding universe we live in when the horizon is inherent   part of the system. Therefore, we opt to present these speculation in the present work.

Non-dispersive contribution with a ``wrong sign" in topological susceptibility (\ref{top1}) obviously implies,  as eq. (\ref{chi})  states, that there is also some energy 
related to this contact term determined  by the surface dynamics  of ``fictitious  particles".
This  $\theta-$ dependent portion of the  energy,
not related to any physical propagating degrees of freedom,  is well established phenomenon and tested on the lattice; it is not part of the debates. What is the part of the debates and speculations 
is the question on how this energy changes when  background varies. In different words, the question we address  in this section can be formulated as follows.  How does   the non-dispersive contribution to the $\theta-$ dependent portion of the  energy vary when conventional Minkowski background is replaced by expanding universe with the horizon size $L\sim H^{-1}$ determined by the Hubble constant $H$\footnote{ Here and in what follows we use parameter $H\sim L^{-1}$ as a typical dimensional factor characterizing the visible size of our universe.  We do not assume at this point  that it is described by FRW metric with a single parameter $H$. In fact, it could be much more generic constructions when the spatial hypersurfaces are embedded in a compact 3d manifold such as, for example,  Bianchi I geometry with few additional parameters. We refer to Appendix B of ref. \cite{dyn} for a short review on this subject in a  given context.}?

The motivation for this question is as follows. 
We adopt the paradigm that the relevant definition of  the energy   which enters the Einstein equations 
 is the difference $\Delta E\equiv (E -E_{\mathrm{Mink}})$, similar to the well known Casimir effect when the observed   energy is in fact a difference 
  between the energy computed for a system with conducting boundaries (positioned at finite distance $d$) and infinite Minkowski space.  In this framework it is quite natural to define the ``renormalized vacuum energy'' to be zero in Minkowski vacuum wherein the Einstein equations are automatically satisfied as the Ricci tensor identically vanishes.   From this definition it is quite obvious that the ``renormalized energy density'' must be proportional to the deviation from Minkowski space-time geometry. 
   This   is  in fact  the standard subtraction procedure  which is normally used for description the horizon's thermodynamics \cite{Hawking:1995fd,Belgiorno:1996yn} as well as  in a course of computations of different Green's function in a curved background by subtracting infinities originated from the flat   space~\cite{Birrell:1982ix}. 
  In the present context  such a definition $\Delta E\equiv (E -E_{\mathrm{Mink}})$ for the vacuum energy for the first time was advocated   in 1967   by Zeldovich~\cite{Zeldovich:1967gd} who argued that  $\rho_{\text{vac}} \sim Gm_p^6 $ with $m_p$ being the proton's mass. Later on such a definition for the relevant energy $\Delta E\equiv (E -E_{\mathrm{Mink}})$ which  enters the Einstein equations has been advocated from   different perspectives in a number of papers,  see e.g.  relatively recent works~\cite{Bjorken:2001pe, Schutzhold:2002pr, Klinkhamer:2007pe, Klinkhamer:2009nn, Thomas:2009uh, Polyakov:2009nq,Krotov:2010ma,Maggiore:2010wr} and references therein.

This is exactly the motivation for question formulated in the previous paragraph:  how does $\Delta E$ scale with $H$? 
The difference $\Delta E$  must obviously vanish when $H\rightarrow 0$ as it corresponds to the transition to flat Minkowski space. How does it vanish? A naive expectation based on common sense suggests that 
$\Delta E \sim \exp(-\Lqcd/H)\sim \exp(-10^{41})$ as QCD has a mass- gap $\sim \Lqcd$, and therefore, $\Delta E$ must not be sensitive to size of our universe $L\sim H^{-1}$. Such a naive expectation formally follows from the dispersion relations similar to (\ref{dispersion})
which dictate that a sensitivity to very large distances must be exponentially suppressed when  the mass gap is present in the system. 

However, as we discussed at length in this paper, along with conventional dispersive contribution we also have the non-dispersive contribution (\ref{top1}, \ref{div1})   which emerges as a result  of topologically nontrivial sectors in four dimensional QCD. This contact term may lead to a power like 
scaling $\Delta E\sim H +{\cal{O}} (H^2)+... $ rather than exponential like $\Delta E \sim \exp(-\Lqcd/H)$ because this term 
(in our framework) is described by massless ghost field (\ref{inverse1})  as discussed in section \ref{QCD}.
The position of this unphysical massless  pole  is topologically protected as eq. (\ref{q}) states, which eventually may result in  power like scaling $\Delta E\sim H +{\cal{O}} (H^2)+... $ rather than exponential like\footnote{If a system is characterized by a single parameter, the curvature, then 
one should expect, on dimensional ground, that the first non-vanishing term in this expansion should be quadratic $\Delta E\sim    H^2$ rather than linear. However, in a generic case one expects a linear non-vanishing term  $\Delta E\sim H +{\cal{O}} (H^2)+... $, see  Appendix B of ref. \cite{dyn} for the details.}.

In fact it was precisely  the  assumption postulated in \cite{dyn,4d} that  the observable Dark Energy (DE) being identified with $\Delta E $ 
  could be  small but not exponentially small.  Similar assumption based on very  different arguments was also advocated in ~\cite{Bjorken:2001pe, Schutzhold:2002pr, Klinkhamer:2007pe, Klinkhamer:2009nn}.
  This postulate on Casimir- like scaling $\Delta E\sim H +{\cal O} (H)^2$  has  recently received a solid theoretical 
  support as  reviewed  below. It is important to emphasize that this term  with power like behaviour  emerges as a  
  result of non-dispersive nature  of topological susceptibility (\ref{top1}), such that no violation of unitarity, gauge invariance or causality occur when the theory is formulated in terms of the unphysical ghosts~\cite{Zhitnitsky:2010ji}.
    If true, the difference between two metrics (expanding universe  and Minkowski space-time) would lead to an estimate 
   \be
   \label{Delta}
   \Delta E\sim H\Lqcd^3\sim (10^{-3} {\text eV})^4,
   \ee
which is amazingly close to the observed DE value today.  It is interesting to note that 
expression (\ref{Delta})  reduces to 
Zeldovich's formula   $\rho_{\text{vac}} \sim Gm_p^6 $ 
   if one replaces $ \Lambda_{QCD} \rightarrow m_p $   and $  H\rightarrow G \Lambda_{QCD}^3$. 
   The last step follows from 
      the  solution of the Friedman equation 
     \be
 \label{friedman}
 H^2=\frac{8\pi G}{3}\left(\rho_{DE}+\rho_M\right),  ~~ \rho_{DE}\sim H\Lqcd^3
  \ee    
  when the DE component dominates the matter component, $ \rho_{DE}\gg\rho_M$. In this case    the evolution of the universe  approaches a  de-Sitter state with constant expansion rate $H\sim G \Lambda_{QCD}^3$ as follows from (\ref{friedman}).

  There are a number of arguments supporting the power like behaviour $\Delta E\sim H +{\cal O} (H)^2$ in gauge theories.  First of all, 
  it is an explicit computation  in exactly solvable  two- dimensional QED  discussed in section \ref{2d} and defined in a box size $L$. The model has all elements crucial for present work: non-dispersive 
  contact term (\ref{exact1}) which emerges  due to the topological sectors of the theory (\ref{chi5}),
  and which can be described using auxiliary  fictitious  ghost fields (\ref{chi6}). This model    is known to be  a  theory of a single physical massive field. Still, 
one can explicitly compute  $\Delta E \sim L^{-1}  $ which is in drastic contrast with  naively expected exponential suppression, $\Delta E\sim e^{-L}$~\cite{Urban:2009wb}. It is important  to emphasize that  this  correction $\Delta E \sim L^{-1}$ while computed  in terms of the ghost's (unphysical) degrees of freedom in our framework, nevertheless represents a gauge invariant physical result. In different words, the final result $\Delta E \sim L^{-1}$ is not related to any violation of gauge invariance though it is computed using auxiliary  fictitious  ghost fields 
similar to computation of the contact term  (\ref{chi6}).

One more support   in  power like behaviour is an  explicit computation in a simple case of  Rindler space-time in four dimensional QCD in the limit when a Rindler observer is moving with acceleration $\Lambda_{QCD}^4\gg a^4\gg m |\<\bar{q}q\>|$ when the interaction term in eq. (\ref{lagC}) can be neglected
~\cite{Zhitnitsky:2010ji, Zhitnitsky:2010zx, ohta}. These computations explicitly show that the power like behaviour emerges in four dimensional gauge systems in spite of the fact that the  physical spectrum is gapped.
\exclude{\footnote{ Formally, in the chiral limit $m_q=0$ there are massless Goldstone bosons, the $\pi$ mesons.  However, one can always consider a case with $N_f=1$  when $\pi$ mesons do not exist,  and a single massive $\eta'$ is present in the system. In this case our statement is mathematically precise: a theory which does not have any massless degrees of freedom as asymptotic states even in the chiral limit, nevertheless demonstrates a power like behaviour in contradiction with naive expectations.}.} In different words, a power like behaviour    is not a specific feature of two dimensional physics as some people (wrongly)   interpret the  results of ref.    \cite{Urban:2009wb}. 

Another argument supporting the  power like corrections is the computation of the contact term in four dimensional QCD defined in a box size $L$. The computations are performed using the so-called instanton liquid model ~\cite{Shuryak:1994rr}.
While the motivation for analysis ~\cite{Shuryak:1994rr} was quite different from our motivation, these model-based computations nevertheless explicitly show 
the emerges of power like corrections to non-dispersive portion of the topological susceptibility. 

Power like behaviour  $\Delta E \sim L^{-1}$  is also supported by  recent lattice results \cite{Holdom:2010ak},
see also earlier paper \cite{Gubarev:2005jm} with  some hints on power like scaling in drastic contrast with naive expectations  $\Delta E\sim \exp-(\Lqcd L)$.   The approach advocated in ref.\cite{Holdom:2010ak}  is based on physical Coulomb gauge  when nontrivial topological structure of the gauge fields is represented by the so-called Gribov copies. It is very different from our approach where we advocate the auxiliary ghost's description to account for this physics.
  Eventually, the physical results must not depend on the different technical tools which are  used   in different frameworks. However, it is not a simple task 
to demonstrate an independence of the results from an employed  technique in a strongly coupled gauge theory!

Finally,  Casimir  like  scaling $\Delta E \sim L^{-1}$ 
can be tested in the so-called ``deformed QCD"  in weakly coupled regime when  all computations are under complete theoretical control \cite{Thomas:2011ee}.
One can explicitly demonstrate that for  the system  defined on a manifold size $    L $ the $\theta$- dependent portion of the energy shows the  Casimir-like
scaling 
 $E = - A\cdot \left[ 1 + \frac{B}{    L } +{\cal O}\left(\frac{1}{    L^2 }\right)  \right]$
 despite  the presence of a mass gap in the system, in contrast with naive expectation 
  $E = - A\cdot \left[ 1 + {B}\exp(-{   L } ) \right]$
  which would normally originate  from any   physical massive propagating  degrees of freedom
  consequent to conventional dispersion relations.    
   
 Another remark worth to be mentioned is that  the sign of $\Delta E\equiv (E -E_{\mathrm{Mink}})$ is always expected to be {\it negative} in conventional quantum field theory computations.  This is due to the fact that some modes can not be accommodated in a system with a nontrivial geometry/boundaries, and therefore the absolute value of $ E_{\mathrm{Mink}} > E$ which corresponds to $\Delta E<0$. The Casimir effect is the well known example when the sign $(-)$ emerges as  a  result  of this subtraction procedure.  The non-dispersive contribution into the energy, on the other hand, 
 being represented by the ghost in our framework will lead to an opposite sign $\Delta E>0$. 
  The positive sign for $\Delta E>0$ is supported by an explicit computations in a simplified settings \cite{Zhitnitsky:2010ji, ohta,Thomas:2011ee}, and is consistent with observations corresponding to the accelerating universe with  $\Delta E>0$. 

 To conclude this section. 
The contact term with ``wrong sign" in topological susceptibility which is   present in gauge theories as a result of a nontrivial topological structure of the theory and which ultimately related to the ``wrong sign" contribution in the entropy computations \cite{Kabat:1995eq,Iellici:1996gv}, as argued in this paper, has another profound consequence. Namely,  the very same physics, and the very same gauge configurations   may lead  to a power like sensitivity from the distant regions  $\Delta E\sim L^{-1}$ in drastic contrast with naive expectations  $\Delta E \sim \exp-(\Lqcd L)$ which should occur from any conventional physical massive propagating  degrees of freedom. If true, one can interpret the extra contribution to the energy (\ref{Delta}), which we identify with DE, as a
result of contact interaction with the horizon. This   interpretation  is consistent with interpretation of the term with a  ``wrong sign" in  entropy computation
\exclude{\footnote{The term with ``wrong sign" in ~\cite{Kabat:1995eq,Iellici:1996gv} represents the difference between the black hole entropy and entropy of entanglement.}}  in two dimensions~\cite{Kabat:1995eq,Iellici:1996gv}, and it is also consistent with our interpretation presented in section \ref{Rindler}, see comments after  formula (\ref{RR}). This interpretation is also consistent with arguments ~\cite{Kabat:1995eq} suggesting that this term corresponds to a contact interaction with horizon in the description of the black hole entropy within a string theory formulation~\cite{Susskind:1994sm}.

   \section*{Conclusion. Future Directions.}\label{conclusion} 
   The main result of this paper is presented in form of a {\it conjecture} formulated at the end of section \ref{2d} and elaborated in section \ref{4d}. Essentially, the basic idea is that the surface term with a ``wrong sign" in  entropy 
computations~\cite{Kabat:1995eq,Iellici:1996gv}  and the  contact term with ``wrong sign" in topological susceptibility  are both originated form the same physics, and both related to the same gauge configurations related to the nontrivial topological structure of the theory.      If this conjecture turns out to be correct, it would unambiguously identify the nature of the well known mismatch between computations of the black hole entropy and entropy of entanglement for vector gauge fields. Similar  mismatch  (but in quite different context)  was  discussed  also in  \cite{Solodukhin:1995ak,Frolov:1996aj,Frolov:1997up}.  In both cases the mismatch is a result of degeneracy of the ground state in the presence of the horizon, see Appendix \ref{subtleties} for the details. 

Another, much more profound consequence is that the same physics which is responsible 
for ``wrong sign" contact term in topological susceptibility, and ``wrong sign" contribution in the entropy computations \cite{Kabat:1995eq,Iellici:1996gv}, may in fact lead to extra vacuum energy (\ref{Delta}) 
(which is identified  with observed DE)
in expanding  universe in comparison with Minkowski space. This extra energy emerges  in gauge theories with   multiple   topological  sectors as a result of merely existence of a causal horizon at distance $L\sim H^{-1}$. Similar phenomenon as we already mentioned occurs in different systems   \cite{Frolov:1996aj,Frolov:1997up} when extra energy 
emerges as a result of  dynamics of the ``soft modes" at the horizon.  The  degeneracy of the vacuum state  in the system discussed in  \cite{Frolov:1996aj,Frolov:1997up} is achieved 
by non-minimally coupling with  a scalar field, see Appendix  \ref{subtleties} for the details, while in our case the presence of topologically distinct sectors  in the system is an inherent  feature of  the QCD dynamics.

Here are some features of these unique  gauge configurations 
which are responsible for ``wrong sign"  in the entropy computations 
and ``wrong sign"  in topological susceptibility and  which are characterized by  very exotic  properties  drastically different from  everything previously known:\\
a) a typical wavelength of fluctuations of the auxiliary ``fictitious particles" is determined by the horizon scale, $\lambda_k \sim 1/H\sim 10\textrm{~Gyr}$, while smaller $\lambda_k \ll1/H$ are exponentially suppressed (\ref{RR}).  Therefore, these modes do not gravitationally clump on distances smaller than the Hubble length,  in contrast with all other types of matter,  and can be identified with the observed properties of DE. Such very large wavelengths prevent us from adopting a meaningful scattering-based description, as the notion of particle is not even defined;\\
b) the corresponding fluctuations are observer dependent, similar to the Unruh radiation, in contrast with any other types of radiation, see section  \ref{Rindler} and also~\cite{Zhitnitsky:2010ji} for detail discussions on problem of measurements in these circumstances;\\
c)  the co-existence of the two drastically different scales ($\Lqcd \sim 100 $ MeV and $H \sim 10^{-33}$ eV) is a direct consequence of the auxiliary conditions~(\ref{gb},\ref{gb-R}) on the physical Hilbert space rather than an \emph{ad hoc} built-in feature such as small coupling or/and extra symmetries  in a Lagrangian.
\exclude{
d) other fine tuning issues such as ``coincidence problems'', ``drastic separation of scales'', ``unnatural weakness of interactions'', etc., which always plague dark energy models, possess a simple and natural explanation within this framework without a single new field/coupling constant in the fundamental Lagrangian of Standard Model, see \cite{dyn} for the details.
In particular, the fine tuning problem which goes under the name of ``cosmic coincidence''  problem finds its natural resolution as follows.
The   vacuum energy which attributed to the ghost in our framework   becomes relevant when its  energy is of the same order of magnitude as $\rho_c$.  Equating these two quantities returns
 $\rho_{DE}\sim H\Lqcd^3 \sim  \rho_c\sim H^2 M_{PL}^2$
 which can be interpreted  as an estimate for the age of the universe when DE becomes the dominating component  in the evolution of the universe.  Numerically, it is estimated as $t_0 \sim H^{-1}\sim M_{PL}^2/\Lqcd^3\sim 10~ $Gyr, which is indeed a  correct estimate for  the lifetime of  the present universe.
}

So, essentially our proposal for the DE can be formulated as follows. The  source for both: DE and    {\it mismatch} 
 between the black hole entropy and entropy of entanglement   is the same and related to the dynamics of   topological sectors of a gauge theory in the presence of the horizon. In other words, the relevant gauge configurations which are responsible for DE are  exactly the same which are responsible for the  ``wrong sign"-contribution  in computations of refs~\cite{Kabat:1995eq,Iellici:1996gv}.  Precisely this contribution represents the mismatch between the black hole entropy and entropy of entanglement. In both cases the source for the extra term is the degeneracy of the vacuum state  which is  represented by different topological sectors, and in our framework is described  by the Veneziano ghost. 
One should also add that  a phenomenological analysis of  the DE model based on this idea and represented by eqs.(\ref{Delta}, \ref{friedman})     has been recently performed in ~\cite{Cai:2010uf}  with conclusion   that this model is consistent with all presently available observational data.

  Another important result of this work can be formulated  as follows. When a system is promoted from Minkowski space to a 
  an expanding  universe, we expect a power like corrections $\Delta E\sim H +{\cal O} (H)^2$ rather than exponentially suppressed corrections, 
  $\Delta E\sim \exp-(\Lqcd/H)$.   This happens   in spite of the fact that the physical Hilbert subspace contains only massive propagating degree  of freedom, and naively the sensitivity to very distant regions should be exponentially suppressed. However, the presence of the non-dispersive contributions (originated from degenerate topological sectors of the theory), which can not be associated with    any physical asymptotic states   {\it falsifies this naive argument}.     Explicit computations in 2d QED  \cite{Urban:2009wb}, in 4d weakly- coupled ``deformed QCD"  \cite{Thomas:2011ee} and  numerical studies in real four dimensional QCD   \cite{Holdom:2010ak} where power like behaviour  $\sim L^{-1}$ is indeed observed, supports our claim.

 What is more remarkable is the fact that some of   fundamental properties  of   gauge theories   discussed in this paper   can be, in principle, experimentally tested in  relativistic heavy ion collider (RHIC) at Brookhaven
 and heavy ion program at LHC.   
 In the ``little bang'' at  RHIC the 
 horizon appears  as a result  of induced acceleration $a\sim \Lqcd$ which itself  emerges as a consequence of high energy  collision. The acceleration   $a\sim \Lqcd$ is a universal number which is determined by strong QCD dynamics, does not depend on energy or other properties of the colliding particles,  and plays the role of   Hubble constant $H\sim 10^{-33} \text{eV}$ of expanding universe,
   see  \cite{Zhitnitsky:2010zx} for the details.
   
\section*{Acknowledgements}

 I am thankful to    participants of the   workshop ``large N gauge theories"  at the Galileo Galilei Institute,   where this work has been presented,  for useful and stimulating discussions. I am also thankful to an anonymous referee for few inspiring questions.  
 This research was supported in part by the Natural Sciences and Engineering Research Council of Canada.
   
 \appendix
  
    \section {Subtleties in definition of energy.}\label{subtleties}
     It is quite natural to expect  that there should be some energy associated with
    fluctuations of fictitious particles (\ref{RR}). As    we argued above these fluctuations, which reflect     the nontrivial topological structure of the gauge theory, is the source for  missmatch   between the black hole entropy and entropy of entanglement in the presence of the horizon.  It is clear that there is some ambiguity in definition of this type of energy due to a number of reasons. First of all, as we discussed in the text the relevant physics is determined by the dynamics on the   surface  rather than in the bulk of the space-time. Therefore, it is not obvious if a simple insertion $\omega_k$ into definition (\ref{H-R}) would properly reflect this feature. Another comment is that the physics of  fluctuations (\ref{RR}) is the observer dependent property similar to the Unruh effect as discussed in section \ref{Rindler}. Therefore, all subtleties related to the Unruh effect are also present here. Also, 
    very large wavelengths of these fluctuations prevent us from adopting a conventional  description in terms of particles, as the notion of particle is not even defined. 
    
    Finally, and most importantly, 
   the  contact interaction which is the main subject of this paper,  can not be expressed in terms of any 
      physical  states, but rather is formulated in terms of fluctuations of fictitious particles with ``wrong" commutation relations. These unphysical states    contribute to the non-dispersive portion of the correlation function, not to the dispersive part which is unambiguously determined by physical spectral function through conventional dispersion relations. Our description in terms of the ghost is simply a convenient way to study this IR physics in a covariant gauge. 
   The same physics  in Coulomb (physical) gauge when the ghost degrees of freedom are not present in the system leads to the long range forces as discussed in simple two dimensional model long ago \cite{KS}. 
       These long range forces prevent distant regions from acting independently. 
   The vacuum in this system is degenerate, and one should consider an infinite superposition of of the winding states $|n\ra$  as originally discussed in \cite{KS}.     We think that precisely this feature prevented  the author of ref. ~\cite{Kabat:1995eq} to use  the physical Coulomb gauge in two dimensions in computations of the entropy, where a covariant gauge has been used instead.  The same  physics in Coulomb gauge in 4d QCD is reflected by existence of the so-called Gribov copies, and one should use some numerical lattice methods to study the relevant physics \cite{Holdom:2010ak}.

   \exclude{ 
    Indeed, 
    our discussions in previous Appendix \ref{2d} have explicitly demonstrated that some correlation functions
    such as topological susceptibility are characterized by two very different contributions: 1) the conventional dispersive portion  which is determined,  through the standard dispersion relations,  by absorptive part  of the amplitude and related to the real physical states; 2) non-dispersive contribution which is not related to any physical states. 
}    
   
    The reason why we pay so much attention to the topological susceptibility $\chi$ and the corresponding contact term $\sim \delta (x)$ which enters the expressions for  topological susceptibility (\ref{exact}) and (\ref{top1})  is due to its relation to the $\theta$ dependent portion of the vacuum energy
  as eq. (\ref{chi}) states. Therefore, the presence of 
    non-dispersive contributions in  $\chi$  
automatically implies presence of the corresponding  non-dispersive contribution in the vacuum energy $E_{\mathrm{vac}}$. At the same time as we discussed in the text  the  non-dispersive contribution in $\chi$ (and in the vacuum energy) can be interpreted  as  a result of nontrivial topological structure of the gauge theories.  These discussions explicitly demonstrate some subtleties on possible  definition of the vacuum energy which should 
 accommodate  the physics related to the contact term discussed in this paper. 
 
    It is quite fortunate that in specific case  with computations of  $\chi$ we can easily separate  non-dispersive and physical  contributions. This is due to the fact, 
    that these two very different terms  contribute   to $\chi$ with opposite signs as discussed in the text. Indeed, 
    the topological susceptibility in pure YM gauge theory (no quarks, and no $\eta'$ contribution)
    according to ~\cite{ven,witten} is given by\footnote{All formulae in this Appendix \ref{subtleties} are written   in Minkowski space
    in contrast with our discussions in the text, where a comparison with   path integral computations and   lattice numerical computations  (which are always   performed in Euclidean space) was made.  In particular, there is factor ``i" in the definition of the correlation function (\ref{top2}).} 
\be
\label{top2}
\chi_{YM}&\equiv& i \lim_{k\rightarrow 0}\int \!\dd^4x e^{ikx} \la T\{q(x), q(0)\}\ra_{W} =  \nonumber \\
&=& -\lambda_{YM}^2< 0, ~~~~~ \lambda_{YM}^2 =\frac{f_{\eta'}^2 m_{\eta'}^2}{4},  
\ee
where $\la...\ra_{W}$ stands for Wick T- product, see below.
The expression on r.h.s of eq. (\ref{top2})   corresponds to the subtraction constant $\sim \delta^4 (x)$ in eq. (\ref{top1}).
This term  has a ``wrong sign" (in comparison with contribution from any real physical states),  the property which motivated the term ``Veneziano ghost''.  
 Indeed, a physical state  of mass $m_G$, momentum $k\rightarrow 0$  and coupling $\la 0| q| G\ra= c_G$ contributes to the topological susceptibility  with the sign which is opposite to (\ref{top2}), 
 \be
\label{G}
&i& \lim_{k\rightarrow 0} \int \!\dd^4x e^{ikx} \la T\{q(x), q(0)\}\ra_{D}  \sim  ~\\ \nonumber
&i&\lim_{k\rightarrow 0}  \la  0 |q|G\ra \frac{i}{k^2-m_G^2}\la G| q| 0\ra \simeq \frac{|c_G|^2}{m_G^2} \geq 0,
\ee
where $\la...\ra_{D}$ stands for Dyson T- product, see below.
 However, the   negative sign for the topological susceptibility (\ref{top2}) is what is required to extract the physical mass for the $\eta'$ meson, see the original reference~\cite{vendiv} for a thorough discussion.  The difference between behaviour (\ref{top2}) and (\ref{G}) is 
 related to inequivalent definitions of these correlation functions. The behaviour (\ref{G}) corresponds to the usual Dyson T-product when only physical states contribute, while eq. (\ref{top2})  corresponds to Wick T-product obtained by variation of the partition function over $\theta$ parameter. The difference in the definitions constitutes  precisely the    subtraction constant $\sim \delta^4 (x)$ in eq. (\ref{top1}). The WI expressed as $\chi_{QCD} (m_q=0)=0$ is  satisfied for the Wick T-product, not for the Dyson T-product.   
 
    It is interesting to note that an analogous  phenomenon (but in quite different  context) was discussed in ref. \cite{Frolov:1996aj,Frolov:1997up}, where it was observed that there are two definitions of energy when  the difference is saturated by the so-called ``soft modes" fluctuating in far infrared at the horizon. First definition is the canonical energy  
    determined by the hamiltonian which is generator of translations of the system along the timelike Killing vector field $\xi^{\mu}$. Since the Killing vector  $\xi^{\mu}$ vanishes at the bifurcation surface of the Killing horizons, the corresponding Hamiltonian is degenerate. Therefore, one can add to the system an arbitrary number of ``soft modes" without changing the canonical energy. These ``soft modes" contribute to the surface integral which was interpreted as a Noether charge of some non-minimally coupled scalar field $\phi_s$. Precisely this  contribution of the ``soft modes"  distinguishes two different 
    definitions of the energy. As argued in ref. \cite{Frolov:1996aj,Frolov:1997up} precisely the dynamics of  ``soft modes"  
    constitutes the difference between Bekenstein -Hawking entropy and the  entropy of entanglement.

    It   is  very similar to our case when 
    we argued that a ``wrong sign" in entropy computations of refs~\cite{Kabat:1995eq,Iellici:1996gv}
    is a result of degeneracy  in the gauge theory represented  in our framework by the fluctuations of the fictitious particles   with a typical  wavelength of order of the horizon scale, $\lambda_k \sim H^{-1}$. In different words, the ``soft modes" from  \cite{Frolov:1996aj,Frolov:1997up} play the same role as pure gauge fields which describe different topological sectors of the theory in our case.  In both cases 
     the effect emerges as a result of  the degeneracy of the ground state in the presence of the horizon, and in both cases 
    the difference is determined by some surface integrals. Further to this analogy, two different  types of  energies which can be reconstructed from  two different definitions of topological susceptibility (\ref{top2}) and (\ref{G})  precisely  correspond to two different types   of energies discussed in refs.  \cite{Frolov:1996aj,Frolov:1997up}.  
   
  The final comment    on the similarities between the two very different systems is as follows. The relevant gauge configurations which are responsible for mismatch between the black hole entropy and entropy of entanglement  from computations~\cite{Kabat:1995eq,Iellici:1996gv} are exactly the same which are responsible for the contact term in  topological susceptibility (\ref{top1}). This mismatch  is    very similar to  extra term discussed  in  \cite{Frolov:1996aj,Frolov:1997up} 
which is  resulted from  the dynamics of the ``soft modes" at the horizon. In both cases the mismatch is a result of degeneracy present in both systems. In refs.   \cite{Frolov:1996aj,Frolov:1997up}  this degeneracy is a  result of non-minimally coupling with  scalar field $\phi_s$. In our cases this degeneracy is a reflection of topological structure of gauge theories, and  inherent feature   of QCD.
However, the outcome of the degeneracy is very similar in both cases as we argued above, and can be interpreted as a contact interaction with horizon, see discussions at the end of section \ref{consequences}.


\end{document}